\documentclass[
twocolumn,
prd,
preprintnumbers,superscriptaddress,showpacs
]{revtex4-1}

\usepackage{graphicx}
\usepackage{amsmath,amssymb,amsfonts}
\usepackage{bm}
\usepackage{color}
\usepackage{comment}
\usepackage{slashed} 
\usepackage{lineno}
\usepackage{url}
\usepackage[colorlinks=true, pdfstartview=FitV, linkcolor=red, citecolor=blue, urlcolor=blue]{hyperref}

\graphicspath{{./plots/}}

\def \be {\begin{equation}}
\def \ee {\end{equation}}
\def \bes {\begin{subequations}}
\def \ees {\end{subequations}}
\def \pd {\partial}
\def \p {\partial}
\def \eq {Eq.~}

\def \<{\langle}
\def \>{\rangle}
\def \+{\dagger}
\def \({\left(}
\def \){\right)}
\def \[{\left[}
\def \]{\right]}
\def \a {\alpha}
\def \b {\beta}

\def \d {\delta}
\def \e {\epsilon}

\def \o {\omega}

\def \l {\lambda}

\def \s {\sigma}

\def \G {\Gamma}

\def \vx {\bm{x}}
\def \vs {\bm{s}}

\def \vb {\bm{b}}
\def \vk {\bm{k}}
\def \vj {\bm{j}}
\def \vA {\bm{A}}

\def \vB {\bm{B}}

\def \vH {\bm{H}}

\def \vE {\bm{E}}

\def \vv {\bm{v}}

\def \he {\hat{e}}
\def \hk {\hat{k}}
\def \hB {\hat{B}}

\def \Im {\rm Im}

\def \Alf {Alfv\'{e}n~}
\def \name  {{\text chiral magnetohelical mode}}
\def \nameA {{\text CMHM}}

\def \CB{{\cal B}}
\def \CO {{\cal O}}

\def \tF {\tilde{F}}

\def \VgSquare {w}

\begin{document}


\title{
Magnetohydrodynamics with chiral anomaly: 
phases of collective excitations and instabilities
}

\author{Koichi~Hattori}
\affiliation{Physics Department and Center for Particle Physics and Field Theory, 
Fudan University, Shanghai 200433, China}

\author{Yuji~Hirono}
\affiliation{Department of Physics,
Brookhaven National Laboratory, Upton, New York 11973-5000, USA}

\author{Ho-Ung~Yee}
\affiliation{Department of Physics, University of Illinois, Chicago, Illinois
 60607, U.S.A.}

\author{Yi Yin}
\affiliation{Center for Theoretical Physics, Massachusetts Institute of Technology, Cambridge, 
MA 02139, USA
}

\date{\today}

\begin{abstract}
We study the relativistic hydrodynamics with chiral anomaly and dynamical
electromagnetic fields, namely Chiral MagnetoHydroDynamics (CMHD). 
We formulate CMHD as a low-energy effective theory based on a generalized derivative expansion. 
We demonstrate that the modification of ordinary
MagnetoHydroDynamics (MHD) due to chiral anomaly can be obtained from
the second law of thermodynamics and is tied to chiral magnetic effect. 
We further study the real-time properties of chiral fluid by solving linearized CMHD equations. 
We discover a remarkable ``transition'' at an intermediate axial chemical potential $\mu_{A}$ between a stable Chiral fluid at low $\mu_{A}$ and an unstable Chiral fluid at large $\mu_{A}$. 
We summarize this transition in a ``phase diagram'' in terms of $\mu_{A}$ and the angle of the wavevector relative to the magnetic field.
In the unstable regime, four collective modes are carrying both magnetic and fluid helicity, in contrary to MHD waves which are unpolarized. 
The half of the helical modes grow exponentially in time, indicating the instability, while the other half become dissipative. 
\end{abstract}

\preprint{MIT-CTP/4958}

\maketitle

%
%
%
%
\section{Introduction}
\label{sec:intro}
Hydrodynamics is a versatile theory describing the real-time dynamics of a given interacting many-body system in long time limit~\cite{landau1987fluid}. 
In this limit most degrees of freedom become irrelevant since they relax at the short time scales. 
The surviving dynamical variables are typically those related to the conservation laws. 
For instance, 
hydrodynamic variables for normal fluid include the energy density $\epsilon$ and the fluid velocity $u^\mu$, 
which correspond to the conservation of energy and momentum respectively. 
A more complicated example is provided by a conducting fluid which is described by magnetohydrodynamics (MHD), 
the theory of which couples the hydrodynamic motion of the fluid to Maxwell's theory of electromagnetism.
The dynamical variables of MHD include not only $\e,u^{\mu}$, but the magnetic field $B^{\mu}$ as well. 
Here,  the field strength tensor and its dual are 
$F_{\mu\nu}$ and $\tF^{\mu\nu}= \frac{1}{2} \e^{\mu\nu\a\b}\,F_{\a\b}=B^{\mu}u^{\nu}-B^{\nu}u^{\mu}+ 
\e^{\mu\nu\a\rho}\,u_{\a} E_{\rho}$, respectively.
The electric charge density $n$ and electric field
$E^{\mu} $ are damped out at a rate proportional to the electric conductivity $\sigma$,
and therefore should not be included as the MHD variables (see Refs.~\cite{Kovtun:2016lfw,Grozdanov:2016tdf,Hernandez:2017mch} and Appendix.~\ref{sec:nvandE}). 
For this reason, we will consider a chiral fluid which is (vector) charge neutral throughout.

The primary purpose of this paper is to study the properties of chiral matter (systems involving chiral fermions) in the long time limit. 
Chiral matter exhibits many interesting phenomena, some of which are closely tied to chiral anomaly. 
We wish to present a hydrodynamic approach for conducting chiral fluid by coupling the dynamics of axial (chiral) charge density $n_{A}$ to MHD, 
and refer the resulting theory as Chiral MHD (CMHD), see Refs~\cite{Yamamoto:2016xtu,Rogachevskii:2017uyc,Boyarsky:2015faa,Giovannini:2016whv} for previous studies. 
This theory would allow us to study those anomaly-induced effects which are absent in ordinary MHD.
The place where such a theory can be potentially applied include 
the quark-gluon plasma (QGP) created by heavy-ion collisions~\cite{Kharzeev:2007jp,Kharzeev:2015znc}, 
newly discovered Dirac and Weyl semimetals~\cite{Li:2014bha,Xiong:2015nna}, 
and the electroweak plasma produced in the primordial universe after the
Big Bang~\cite{Joyce:1997uy,PhysRevLett.108.031301}.

However, $n_{A}$ is distinguished from standard MHD variables because of their difference in the long wavelength behavior of the relaxation rate. 
Specifically, $\G_{A}$, the relaxation rate of $n_{A}$, is finite in small wavevector (or gradient) $k$ limit since axial current $J^{\mu}_{A}$ is not conserved due to quantum anomaly. 
Therefore one has to identify additional small parameter to make $\G_{A}$ parametrically small so that $n_{A}$ can be counted as a parametrically slow mode (see Ref.~\cite{hydroplus} for a discussion on the extension of hydrodynamics with parametrically slow modes in generic situations.)
We identify this additional small parameter as the anomaly coefficient $C_{A}$ (see \eq\eqref{CA-QED} below), and will work in the limit $C_{A}\ll 1$. 
This identification is natural because $C_{A}$ tracks the effects of quantum anomaly,
which are typically suppressed by additional power of $\hbar$.
Without such anomalous effects,  $n_{A}$ would be conserved and hence $\G_{A}$ would vanish in $C_{A}\to 0$ limit. 
In fact, we shall see $\G_{A}\propto C^{2}_{A}$, 
which is in analogous to the relaxation rate of standard MHD variables which is proportional to $k^{2}$.

The very presence of this additional small parameter $C_{A}$ also necessitates the generalization of the standard procedure of derivative expansion for hydrodynamics to construct CMHD. 
We will formulate CMHD based on the double expansion in terms of the number of gradient $k$ (times mean free path $l_{{\rm map}}$) and $C_{A}$ in Sec.~\ref{sec:CMHD}.
By double expansion, we mean both $C_{A}$ and $k\, l_{\rm mfp}$ are small, but we do not assume any hierarchy between them.
Here and hereafter, 
let us use $\CO(\delta)$ to denote terms of the order $\CO(k)$ ($k$ times mean free path) and/or $\CO(C_{A})$.
As detailed below, 
we will express the stress-energy tensor $T^{\mu\nu}$, $J^{\mu}_{A}$ and $E^{\mu}$ in terms of CMHD variables $\e, u^{\mu}, B^{\mu},n_{A}$ up to first order in $\CO\(\d\)$. 
From now on, quantities at $\CO(\d)$ are sometimes labeled with subscript $(1)$.
In our derivation, the stringent constraint imposed by the second law of thermodynamics is taken into account.  
The result of doing so yields
\begin{eqnarray}
\label{E-relation}
E^{\mu}_{(1)}&=&
- \frac{1}{\s}\, 
\[\, C_{\rm A}\, \mu_{\rm A}\, B^{\mu}+\b^{-1}\e^{\mu\nu\a\rho}\, u_{\nu} \,\pd_{\alpha}\(\beta\,H_{\rho}\)\, \] \, . 
\end{eqnarray}
In Eq.~\eqref{E-relation} $\b$ is the inverse of the temperature and $H^{\mu}$ is the in-medium magnetic field. 
While $\CO(k)$ term in Eq.~\eqref{E-relation} already shows up at MHD, 
the presence of $\CO(C_{A})$ term there, being proportional to axial chemical potential $\mu_{A}$, is the distinctive feature of CMHD. 
We shall see this term is closely related to the chiral magnetic effect (CME)~\cite{Nielsen1983389,Vilenkin:1980fu,Fukushima:2008xe} (see Refs.~\cite{Kharzeev:2012ph,Kharzeev:2013ffa,Miransky:2015ava,Kharzeev:2015znc,Hattori:2016emy} for a recent review), 
the generation of the electric current by a magnetic field in chiral matter with non-zero $n_{A}$.

We next explore the real-time properties of a Chiral fluid by solving linearized CMHD,
and study the corresponding collective excitations in Sec.~\ref{sec:dispersion}. 
Our chief observation is the ``transition'' at an intermediate $\mu_{A}$ between a stable Chiral fluid at low $\mu_{A}$ and an unstable Chiral fluid at large $\mu_{A}$. 
In the unstable regime, there will be four collective modes carrying both magnetic and fluid helicity, in contrary to waves in MHD which are unpolarized. 
Half of the helical modes have positive imaginary, indicating the instability. 
The formulation of CMHD based on a new derivative expansion scheme together with the discovery of a qualitative difference in the dynamical properties of Chiral fluid are the main findings of this paper.

\section{Chiral MHD}
\label{sec:CMHD} 

The equations of motion for CMHD variables $\e, u^\mu, B^\mu,n_{\rm A}$
consist of the energy-momentum conservation, the Bianchi identity and the anomaly equation (we use the most minus sign convention for the metric $g^{\mu\nu}$):
\bes
\label{MHD-eq}
\begin{eqnarray}
\label{energy-momentum-conservation}
\pd_{\mu}\, T^{\mu\nu} &=& 0\, ,
\\
\label{Bianchi}
\pd_{\mu}\, \tF^{\mu\nu} &=& 0\, ,
\\
\label{anomaly}
\pd_{\mu}\, J^{\mu}_{\rm A} &=& - C_{\rm A}\, E\cdot B  \, , 
\end{eqnarray}
\ees
where $J^{\mu}_{\rm A}$ is the axial current, and anomaly coefficient is given by:
\begin{equation}
\label{CA-QED}
C_{A}= \frac{e^{2}}{2\pi^{2}}\, .
\end{equation}
Here, $T^{\mu\nu}$ is the energy-momentum tensor of the total system (i.e., the fluid and EM fields) so that it is conserved. 
\eq\eqref{anomaly} shows that the evolution of $n_{A}$ is closely related to the dynamics of EM fields. 
Indeed, the reconnection of magnetic flux could induce the change of $n_{A}$~\cite{Hirono:2016jps}. 
Depending on the microscopic details of the systems under study, there could be other processes which contribute to the relaxation of $n_{A}$. 
For example, topological sphaleron transitions can also change the axial charge in a quark-gluon plasma (QGP), but the transition rate scales with $g^{10}$ for weakly coupled QGP.  
We have neglected those additional processes when writing down \eq\eqref{anomaly} to simplify the discussion, and leave the extension to future work. 
%

At zeroth order $\CO(\delta^{0})$,
the form of $T^{\mu\nu}, \tF^{\mu\nu}$ is identical to those of MHD.  
Therefore
\bes
\label{eq:constitutive}
\begin{eqnarray}
\label{Tmunu}
T^{\mu\nu}&=&\e\, u^{\mu}u^{\nu} - p
\(g^{\mu\nu} -u^\mu u^\nu\) - H^{\mu}B^{\nu}
\nonumber \\
&& 
+ T_{(1)}^{\mu\nu}\,+\CO(\d^{2})\,  , 
\\
\label{Ftilde}
\tF^{\mu\nu} &=&B^{\mu}u^{\nu}-B^{\nu}u^{\mu}+ 
\e^{\mu\nu\a\rho}\,u_{\a} E_{\rho (1)}\,+\CO(\d^{2})\, ;
\\
\label{Jmu}
J^{\mu}_{\rm A} &=& n_{\rm A}\, u^{\mu}+ J_{\rm A(1)}^{\mu}\,+\CO(\d^{2})\, .
\end{eqnarray}
\ees
see also Appendix.~\ref{sec:cmhd} for a detailed discussion. 
Here, the in-medium magnetic field $H^{\mu}$ is conjugate to $B_{\mu}$, i.e., 
\begin{eqnarray}
\label{ds}
d\,s = \b d\e -\(\b\mu_{\rm A}\)\, dn_{\rm A} +\b H_{\mu}\, d\, B^{\mu}\, .
\end{eqnarray}
The entropy density $s$ and pressure $p$ 
are related by the thermodynamic relation: 
\begin{eqnarray}
\label{p}
p= -\e + \mu_{\rm A}\, n_{\rm A}+\b^{-1}\, s\, -H\cdot B\, . 
\end{eqnarray}
We remind the reader that $T^{\mu\nu}$ refers to the total stress tensor (i.e., the sum of fluid and Maxwell stress tensor) of the system.
Therefore there is a $B^{\mu} H^{\nu}$ term in Eq.~\eqref{Tmunu}, and a dependence of $p$ on $B$. 
They are originated from the Lorentz force that the charged fluid would experience (c.f.~Ref.~\cite{Israel1978}).

We next consider the entropy current $S^{\mu}=s\,u^{\mu}+s_{(1)}^{\mu}$ and require the positivity of the entropy production $\pd_{\mu}S^{\mu} \geq 0$.
Transforming $\beta u_{\nu}\pd_{\mu}\,T^{\mu\nu}+\b \mu_{A}\,\pd_{\mu}\, J^{\mu}+\b H_{\mu}\pd_{\nu} \tilde{F}^{\mu\nu}$ using Eq.~\eqref{MHD-eq}, 
we find (in Landau fluid frame $u_{\mu}\, T^{\mu\nu}_{(1)}=0$):
\begin{eqnarray}
\label{entropy-1}
&\,&
\beta \(\pd_{\mu} u_{\nu}\)\,T^{\mu\nu}_{(1)}
+\[-\pd_{\mu}\(\b\mu_{A}\)\, J^{\mu}_{A(1)}\] +
\nonumber\\
&\,&+ E_{\mu(1)}\,\[\,
  C_{\rm A}\, \mu_{\rm A}\, B^{\mu}
 + \b^{-1} \e^{\mu\nu\a\rho}\, u_{\nu} \,\pd_{\alpha}\(\beta\,H_{\rho}\) \]\geq 0
\end{eqnarray}
with $S_{(1)}$ given by \eqref{s1-result},
see Appendix.~\ref{sec:cmhd} for more details. 
We note $\pd_{\mu}\(s u^{\mu}\)=0$ (c.f~Appendix.~\ref{sec:ideal}), meaning there is no entropy production for zeroth order (i.e. ideal) CMHD. 
In order to satisfy the condition \eqref{entropy-1}, it is sufficient to require each of the three terms on the L.H.S of Eq.~\eqref{entropy-1} to be positive definite (c.f.~\eq\eqref{condition-detail}). 
The expression for $T^{\mu\nu}_{(1)}$ in MHD has been determined previously~\cite{Huang:2009ue,Huang:2011dc, Kovtun:2016lfw,Grozdanov:2016tdf,Hernandez:2017mch}, 
and satisfies $\(\pd_{\mu}u_{\nu}\)\,T^{\mu\nu}_{(1)}\geq 0$. 
In another word, the constitutive relation for $T^{\mu\nu}_{(1)}$ is identical to that of MHD. 
Meanwhile, $\[-\pd_{\mu}\(\b\mu_{A}\)\, J^{\mu}_{A(1)}\] \geq 0$ will be satisfied if $J^{\mu}_{A}=\l_{A}\pd^{\mu}(\b\mu_{A})$, where $\l_{A}$ is a positive transport coefficient. 
We finally turn to the condition:
\begin{eqnarray}
\label{E-condition}
E_{\mu(1)}\,\[\,
  C_{\rm A}\, \mu_{\rm A}\, B^{\mu}
 + \b^{-1} \e^{\mu\nu\a\rho}\, u_{\nu} \,\pd_{\alpha}\(\beta\,H_{\rho}\) \]\geq 0\, , 
\end{eqnarray}
which requires 
 $E^{\mu}$ to be of the form given by Eq.~\eqref{E-relation}.
 Here, a positive constant $\sigma$ will be identified with the electric conductivity shortly.
For simplicity, we assume that $\s$ is isotropic, which is the case in a weak magnetic field limit. 
Nevertheless, our conclusion on the CME current 
holds even with a general tensor structure of $\s$, as is shown in in Appendix~\ref{sec:first-order}.

We now demonstrate that the first term on the R.H.S. of Eq.~\eqref{E-relation} is tied to CME. 
To simplify our analysis, 
we take $\b$ to be homogeneous and rewrite the spatial part of Eq.~\eqref{E-relation} in the local rest frame of fluid:
\begin{eqnarray}
\label{E-rest}
\vE=\s^{-1}\(-C_{A}\mu_{A} \vB +\nabla\times \vB\).
\end{eqnarray}
Employing Ampere's law $\vj=\nabla\times \vB$, 
we have:
\begin{eqnarray}
\label{jv}
\label{jv}
\vj= C_{A}\mu_{A} \vB +\sigma \vE\, . 
\end{eqnarray}

Two implications follow from Eq.~\eqref{jv}. 
First, $\s$ in Eq.~\eqref{E-relation} has to be the conductivity so that $\sigma \vE$ is the usual Ohm current. 
Second, we now recover the CME current from Eq.~\eqref{E-relation}.
Notice that the dynamical variables as well as counting scheme here are different from earlier works. 
For example, 
$E^{\mu},B^{\mu}$ are non-dynamical, and are counted as $\CO(k)$ in Ref.~\cite{Son:2009tf}. 
Despite of these differences, we find the same form of the CME
current, which exemplifies the universal nature of CME.
To best of our knowledge,
the demonstration of this universality within the framework of CMHD based on the second law of thermodynamic is new in literature.
In fact, there are studies which show that the form of other anomaly-induced effects, in particular that of CVE, is non-universal with dynamical gauge field~\cite{Hou:2012xg}. 
In this regard, our result on the universal form of CME contribution is quite remarkable. 

We wish to reiterate the novelty and necessity of double expansion in terms of $C_{A}$ and gradient in CHMD, through which we obtain \eq\eqref{E-relation} and hence the manifestation of CME. 
If we were using the conventional gradient expansion, then the first term, i.e. CME term, in \eq\eqref{E-relation} has to be counted as zeroth order in gradient. 
This contradicts with the assumption that electric field $E^{\mu}$ is not a slow variable and can not be counted as the zeroth order term in the expansion. 
Generally speaking, the very existence of non-hydrodynamic slow modes suggests the presence of at least one additional small parameter which controls the slowness of such modes in the system of interest. 
For the case of chiral fluid with slow evolving $n_{A}$, we identify this slow parameter as $C_{A}$.

\section{Collective modes}
\label{sec:dispersion}

%
%
%
\begin{figure}
\vspace{-0.5cm}
\begin{center}
\includegraphics[scale=.6]{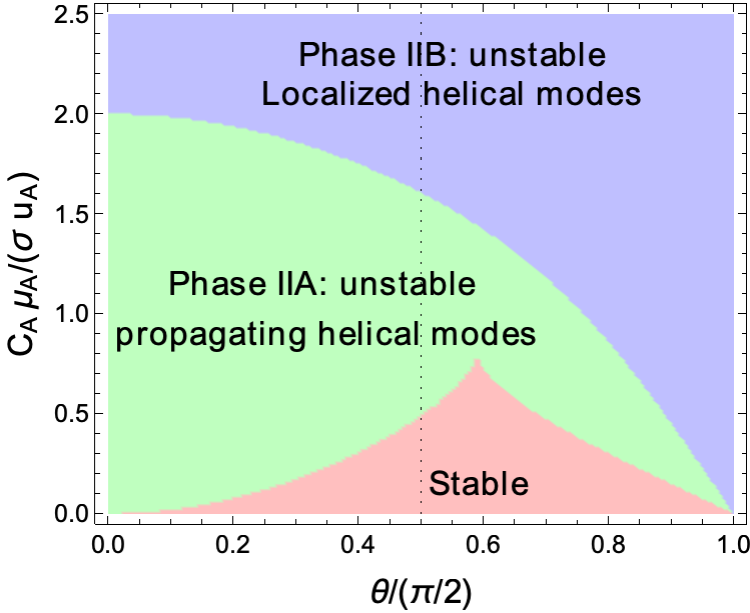}
\end{center}
\vspace{-0.5cm}
\caption{
(Color online) A ``phase diagram'' charting the stable and unstable regimes of chiral fluid in the $\theta$ - $\e_{\rm A}$ plane (with $c_{\rm
 s}/u_{\rm A}=0.6$), see text.
Dashed vertical (horizontal) curve represents the fixed value of $\theta$ ($\e_{\rm A}$) used in Fig.~\ref{fig:ModesvsnA} 
}
\label{fig:ModeDiagramCold}
\vspace{-0.5cm}
\end{figure}
%
%
%

%
%
\begin{figure*}[t!]
\vspace{-1cm}
\begin{center}
\includegraphics[scale=0.4]{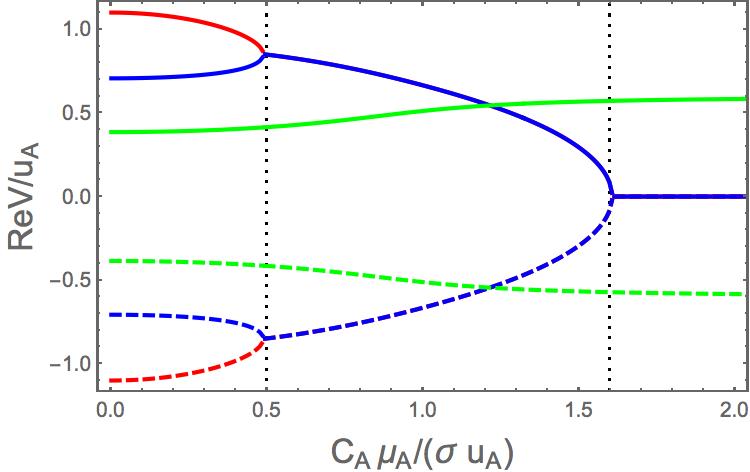} (a) 
\includegraphics[scale=0.4]{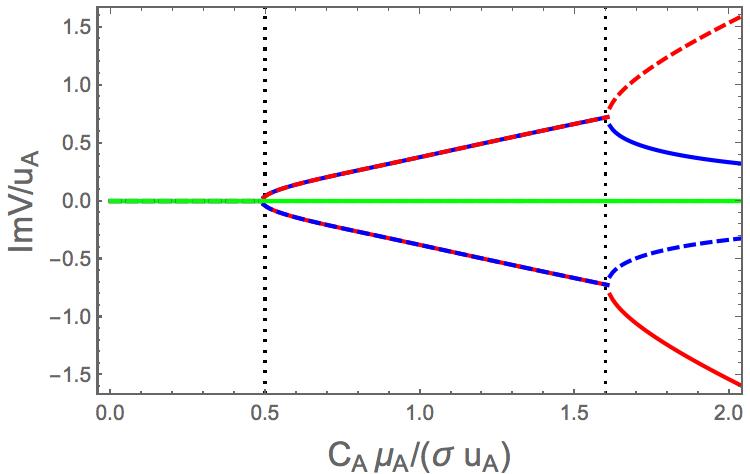} (b) 
\includegraphics[scale=0.43]{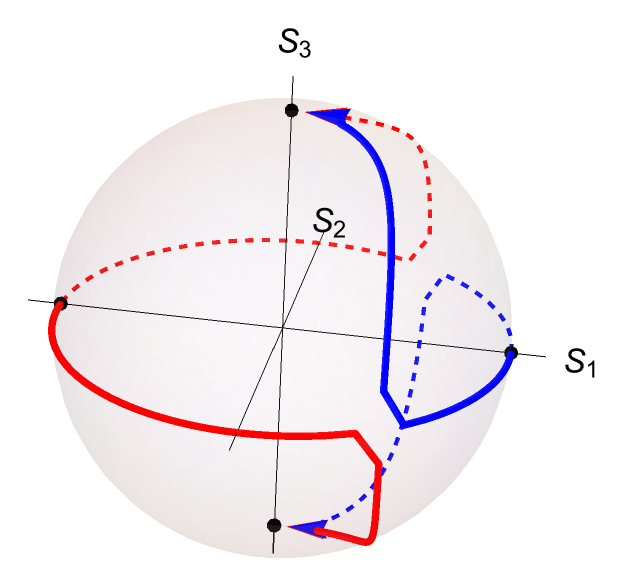} (c) 
\end{center}
\vspace{-0.5cm}
\caption{
\label{fig:ModesvsnA}
Plots of (a) the real part and (b) the imaginary part of $ V $, 
and (c) the polarizations vs $\e_{\rm A}/u_{\rm A}$ at $\theta=\pi/4,
c_{\rm s}/u_{\rm A}=0.6$, see text for more.
The corresponding modes are shown in the same colors through (a) to (c). 
Vertical lines show the phase boundaries in Fig.~\ref{fig:ModeDiagramCold}. 
A pair of modes corresponding to the green beaches in (a) and (b) always stay at the equator and are not shown in (c). 
}
\vspace{-0.2cm}
\end{figure*}
%
%

We now consider fluctuations around a uniform static background:
$\e(t,\vx)=\e +\delta \e(t,\vx)$, $u^{\mu}(t,\vx)=\(1,\vv(t,\vx) \)$, $B^{\mu}(t,\vx)=\(-\vB\cdot \vv(t,\vx),-\vB+\d\vB(t,\vx)\)$ and $n_{A}(t,\vx)=n_{A}+\delta n_{A}(t,\vx)$.
Our formulation of CMHD is valid as far as both $C_{A}$ and the gradient are small. 
In what follows, we will solve linearized CMHD for those fluctuations in frequency $\o$ and wavevector $\vk$ space by specifying the following hierarchy among $k$ and other scales. 
First, our focus will be on the regime  $ k\ll C_{A}\mu_{A}$, 
since early studies indicate that chiral plasma would become unstable in this momentum regime. 
Consequent, we neglect the second term on R.H,S of \eqref{E-relation}. 
In this limit, $\pd_{t}n_{A}=C_{A}\vE\cdot \vB$ will become a relaxation equation of $n_{A}$ upon substituting $\vE=-C_{A}\mu_{A} \vB/\sigma$. 
The corresponding relaxation rate $\G_{A}\sim C^{2}_{A}$, as we advertised earlier. 
Next,  we will require $\o(k) \gg \G_{A}\sim C^{2}_{A}$ so that the evolution of $\delta n_{A}(t,\vx)$ is decoupled from other fluctuating variables. 
In small $C_{A}$ limit, there is indeed a wide range of $k$ satisfying both conditions. 
Finally, we assume $\eta k,\zeta k \ll \e$ as in ordinary hydrodynamics, where $\eta$ and $\zeta$ are shear and bulk viscosity respectively, 
For this reason, we will not include the contribution due to $T^{\mu\nu}_{(1)}$ from now on.

To proceed, we will use a simplified equation of state $p=p_{f}(\e_{f})- \CB^{2}/2$ so that $H^{\mu}= B^{\mu}$.
Here $\e_{f}=\e - \CB^{2}/2$ and $p_{f}$ are the fluid part of the energy density and pressure respectively, where we have defined
\begin{eqnarray}
\label{B-magnitude}
\CB\equiv \sqrt{-B\cdot B}\, . 
\end{eqnarray}
For definiteness, we will use $p_{f}(\e_{f})=c^{2}_{s}\, \e_{f}$ where $c_{s}$ is the sound velocity. 
The linearized equations for (rescaled) fluctuation fields
\begin{eqnarray}
\d\tilde{\e}_{f}\equiv \frac{\d \e_{f}}{\(e_{f}+p_{f}\)}\, , 
\qquad
\vb \equiv \frac{\d\vB}{\sqrt{\e_{f}+p_{f}}}\, 
\end{eqnarray}
 now reads, 
\bes
\label{CMHD-lin}
\begin{eqnarray}
\label{ef}
i\o \d \tilde{\e}_{f}&=& - i k v_{L}\, ,
\\
\label{vL}
i\o v_{L}&=& - i  k \( c^{2}_{s}\,\d \tilde{\e}_{f}-u_{A}\,\sin\theta\, b_{2}\) ,
\\
\label{Lorentz}
i\o v_{1}&=&i \, k u_{A}\cos\theta\, b_{1}\, ,
\quad
i\o v_{2}= i \, k u_{A}\cos\theta\, b_{2}\, ,
\\
\label{b1}
i\o \,b_{1}&=& i  k \( u_{A}\cos\theta v_{1} +\e_{A}\,b_{2}\)\, ,
\\
\label{b2}
i\o b_{2}&=&
i k\[u_{A}\(\sin\theta v_{L}+\cos\theta v_{2}\) - \e_{A}\,b_{1}\]
 \, ,
\end{eqnarray}
\ees
where $\theta$ is the relative angle between $\vk$ and $\vB$. 
By introducing a standard orthogonal unit basis $\he_{1}\propto \hB\times \hk, \hat e_{2}=\hk\times \hat e_{1}$ and $\hk$~\cite{jackson1975classical},
we have decomposed the fluctuation fields as $\vv=v_{1}\, \he_{1}+v_{2}\, \he_{2}+v_{L}\, \hk$ and $\vb=b_{1}\, \he_{1}+b_{2}\, \he_{2} $ (note that $\nabla \cdot\vb=0$).
For later convenience, we have introduced two important dimensionless parameters, namely,  \Alf velocity $u_{\rm A}\equiv B/\sqrt{\e_{f}+p_{f}}$ and $\e_{A}=C_{A}\mu_{A}/\sigma$. 
We have further assumed $u_{A}\ll 1$, 
and have dropped terms suppressed $u^{2}_{\rm A}\ll 1$ when writing down Eq.~\eqref{CMHD-lin}.
However, $\e_{\rm A}/u_{\rm A}$ can be ${\cal O}(1)$ since $\e_{{\rm A}}\propto C_{A}\ll 1$.

The dispersion relation $\o(k) =V k$ of collective modes can be determined by solving Eq.~\eqref{CMHD-lin}, where  ``group velocity'' $V$ satisfies:
\begin{eqnarray}
\label{VG-eq}
&&
\( V^{2}- u^{2}_{A}\cos^{2}\theta\)
\left[ V^{4} -\( u^{2}_{A}+c^{2}_{\rm s}\)
 V^{2}+c^{2}_{\rm s}\,u^{2}_{\rm A}\,\cos^{2}\theta \right]
\nonumber \\
&\,& 
\hspace{0.2cm}
+\e^{2}_{\rm A} \, V^{2}\(\, V^{2}- c^{2}_{\rm s}\,\) =0 \, .
\end{eqnarray}
Eq.~\eqref{VG-eq} has six roots, corresponding to six collective modes.
Furthermore,
by expressing $v_{1,2}$ in terms of $b_{1,2}$ using Eq.~\eqref{Lorentz}, 
and substituting the resulting expressions  into Eq.~\eqref{b1}, 
we found
\begin{eqnarray}
\label{polarization-V}
 \frac{v_{1}}{v_{2}} &=& 
 \frac{b_{1}}{b_{2}} = \frac{\e_{\rm A} \, V}{u^2_{\rm A}\cos^{2}\theta - V^2}\, . 
\end{eqnarray} 
Eq.~\eqref{polarization-V} is very informative in at least two aspects. 
First, it implies that the relative phase between $b_{1}, b_{2}$ is the same as that of $v_{1}, v_{2}$. 
If a collective mode carries the positive (negative) magnetic helicity, it also carries the fluid helicity of the same chirality. 
Such mode will be called RH (LH) mode below.
By definition, magnetic and fluid helicity are positive (negative) if  $\d\vA\cdot \d\vB >0 (<0)$ and $(\nabla\times \vv)\cdot \vv>0 (<0) $ respectively.  
Here $\d\vA$ is the vector gauge potential satisfying $\vB	=\nabla\times\d\vA$. 
Second,
Eq.~\eqref{polarization-V} will tell us the polarization of each collective mode with a given $V$.
For instance, Eq.~\eqref{polarization-V} implies that a mode with a real-valued $V$ is linearly polarized whereas that with a purely imaginary $V$ is circularly polarized.

Let us first consider Eq.~\eqref{CMHD-lin} in two limiting cases. 
In the limit $\e_{\rm A}/u_{A}=0$,  Eq.~\eqref{CMHD-lin} is reduced to the linearized MHD.
The collective modes are well-known as the \Alf wave and the fast and slow magnetosonic waves \cite{biskamp1997nonlinear} with group velocity given by $V_{A}, V_{F}, V_{S}$ respectively.
Solving Eq.~\eqref{VG-eq} at $\e_{A}=0$, one finds:
\begin{eqnarray}
\label{VA}
V^{2}_{A}
&=&u^{2}_{A}\cos^{2}\theta\, , 
\\
\label{VFS}
V^{2}_{F,S}
&=& \frac{\(u^{2}_{A}+c^{2}_{s}\)\pm \sqrt{u^{4}_{A}+c^{4}_{s}-2u^{2}_{A}\,c^{2}_{s}\cos\(2\theta\)}}{2}
\end{eqnarray}
where $+$ ($-$) sign corresponds to $V_{F} (V_{S})$.
We note $V_{F,S}, V_{A}$ are \textit{real}, indicating MHD waves are \textit{unpolarized} and ordinary MHD systems are stable. 

Now we turn to the opposite limit $\e_{\rm A}/u_{A}\gg 1$. 
In this case,  
the evolution of $b_{1},b_{2}$ is decoupled from that of $\vv, \delta \tilde{e}_{f}$, 
and is described by setting $u_{A}=0$ in Eqs.~\eqref{b1}, \eqref{b2}.
The corresponding collective modes are $\(b_{1}, b_{2}\)\propto \(1, \pm i \)$, 
corresponding to \textit{circularly polarized} (helical) magnetic fields, 
with \textit{purely imaginary} $V= \pm i \e_{A}$. 
Such modes are refered in literature as Chern-Simon (CS) modes~\cite{Laine:2005bt}. 
Half of CS modes have a positive imaginary part, signifying the instability of a Chiral plasma, as discussed earlier in Refs~\cite{Laine:2005bt,PhysRevLett.108.031301,Akamatsu:2013pjd}.
%
%

If we put the preceding analysis at small and large $\mu_{A}$($\e_{A}$) together, 
we conclude that Chiral fluid with dynamical magnetic field must have at least one ``transition'' at an intermediate $\mu_{A}$ between a stable Chiral fluid at low $\mu_{A}$ and a unstable Chiral fluid at large $\mu_{A}$. 
We now put this qualitative expectation on a quantitative basis by computing $V$ of each collective mode at given $ \e_{\rm A}/u_{A}$ and $\theta$ by solving Eq.~\eqref{VG-eq}.
It is sufficient to consider $0\leq\theta\leq \pi/2$ since $V$ determined from Eq.~\eqref{VG-eq} will depend on $\cos^{2}(\theta)$ only. 
In Fig.~\ref{fig:ModeDiagramCold}, we present a ``phase diagram'' 
which charts stable and unstable regimes (``phases'') in $\e_{\rm A}-\theta$ plane. 
%

In low $\mu_{A}$ ``phase'', 
i.e. ``Phase I'' (red regime), Chiral fluid is stable. 
All modes are akin to MHD waves up to the modifications of the group velocities which are real-valued. 
However, 
as we increase $\e_{A}$, the system transits to the unstable phase, i.e. ``Phase II''. 
While there are still two modes similar to ordinary MHD waves in this phase, 
the remaining four modes become helical and have non-zero ${\rm Im} V$.
The half of those modes have ${\rm Im}V>0$, indicating the instability. 
As a specific example, 
Fig.~\ref{fig:ModesvsnA}~(a, b) show the real and imaginary parts of $V$ for all six modes as functions of $\epsilon_{\rm A}$ at a fixed $\theta=\pi/4$.

Fig.~\ref{fig:ModeDiagramCold} reveals that chiral fluid is stable for a small but finite $\mu_{A}$ for any generic $\theta$. 
This is in stark contrast with the case of chiral plasma which would become unstable in the presence of an infinitesimal small $\mu_{A}$~\cite{Laine:2005bt,PhysRevLett.108.031301,Akamatsu:2013pjd}. 
To understand this difference, 
let us consider, without losing generality, the modification of $V$ at small $\e_{A}$ of the mode which corresponds to \Alf wave at $\e_{A}=0$ . 
Substituting $V=V_{A}+\Delta V$ into Eq.~\eqref{VG-eq} and expanding it to linear order in $\Delta V$, 
we find:
\begin{eqnarray}
\label{V-per}
\Delta V=-\frac{\e^{2}_{A}\,\(V^{2}_{A}-c^{2}_{s}\)V_{A}}{2\(V^{2}_{A}-V^{2}_{F}\)\(V^{2}_{A}-V^{2}_{S}\)}\, . 
\end{eqnarray}
To obtain Eq.~\eqref{V-per}, we have used the fact that the first line of Eq.~\eqref{VG-eq} can be put into to the form $\(V^{2}-V^{2}_{F}\)\(V^{2}-V^{2}_{S}\)\(V^{2}-V^{2}_{A}\)$. 
Eq.~\eqref{V-per} shows that $\Delta V$ remains real, i.e. $\Im V=0$, for sufficiently small $\e_{A}$ and for $\theta$ at which there is no degeneracy among $V_{F,S},V_{A}$ so that the denominator of Eq.~\eqref{V-per} is nonzero. 
Consequently, chiral fluid is stable in this case except for $\theta=0$ that $V_{F}=V_{A}=u_{A}$ and for $\theta=\pi/2$ that $V_{S}=V_{A}=0$ (c.f.~Eqs.~\eqref{VA}, \eqref{VFS}). 
Putting it in a perhaps more intuitive, albeit less rigorous way,
we can think of those helical modes in the unstable phase as originated from a mixture of different linearly polarized MHD waves. 
Such mixture will not be energetically favorable unless group velocity of two different MHD waves become identical.

We now focus on those four modes which become helical in the unstable phase.
One can show from Eq.~\eqref{polarization-V} that those helical modes have the remarkable properties of ``selective growth'', 
namely,
when $\e_{\rm A}>0$ ($\e_{\rm A}<0$), 
the $Im V$ of RH(LH) modes is positive, meaning that RH (LH) modes are ``selected'' to grow exponentially in time. 
CS modes for chiral plasma also exhibits the properties of ``selective growth'', and its physical origin has been discussed in many early works (e.g. Refs.~\cite{Laine:2005bt,PhysRevLett.108.031301,Akamatsu:2013pjd,Hirono:2015rla}). 
Since chiral anomaly can re-distribute helicity between the fermionic and magnetic parts, the chiral plasma will tend to minimize the energy cost at a fixed helicity by populating modes with a definite helicity. 
The physics is similar here. 
What is distinctive about helical modes of Chiral fluid is that they carry non-zero fluid helicity in addition to magnetic helicity, since magnetic field and fluid field are coupled to each other.

To illustrate this close relationship between the chirality of the helicity and the instability of those collective modes, 
we plot the trajectories of the Stokes vector~\cite{jackson1975classical} 
$\vs
=\(b^{2}_{1}-b^{2}_{2},2{\rm Re}\[ b_{1}\, b^{*}_{2}\] ,2\, {\rm Im}\[ b_{1}\, b^{*}_{2}\] \)/\(b^{2}_{1}+b^{2}_{2}\)
$ 
corresponding to those four modes
with varying $\e_{A}$ at $\theta=\pi/2$ on a unit sphere (the Poincar\'e sphere) in Fig.~\ref{fig:ModesvsnA}~(c). 
By definition, 
a point on the equator of the  Poincar\'e sphere specifies a linear polarization, 
while that on the upper and lower hemispheres are 
left-handed and right-handed polarization, respectively. 
In particular, the north and south poles correspond to the circular polarizations.
The red and blue trajectories start at $\vs=\(-1,0,0\)$ and $\vs=\(1,0,0\)$ (at the equator of the Poincar\'e sphere) respectively, corresponding to $\e_{A}=0$, 
and ``flow'' to the upper/lower hemispheres when the transition occurs from Phase I to II, and eventually approach the north/south poles. 
Notice the correspondences between the polarization states 
and the dispersion relations shown in the same colors through Fig.~\ref{fig:ModesvsnA} (a) to (c). 

%
%
%

While in high $\mu_{A}$ ``phase'', i.e. ``Phase IIB'', 
the values of $V$ of helical CMHD modes are purely imaginary, similar to CS modes in the Chiral plasma, 
those four helical modes have complex-valued $V$ at an intermediate $\mu_{A}$ regime (``phase IIA''). 
We will call them ``\name \ (\nameA).'' 
It might be useful to view \nameA s~as an outcome of an interesting hybridization of MHD waves and CS modes.
They ``inherit'' the ability of propagation in space ($ReV\neq 0$) from MHD waves, 
and that of carrying magnetic helicity from CS modes. 
The presence of such new collective modes have not been reported in the preceding studies of CMHD~\cite{Yamamoto:2016xtu,Rogachevskii:2017uyc,Boyarsky:2015faa,Giovannini:2016whv}.
\nameA s~are also distinct from collective modes in chiral fluid with non-dynamical magnetic field~\cite{Newman:2005hd,Kharzeev:2010gd,Jiang:2015cva,Yamamoto:2015ria}. 
The emergence of \nameA s clearly demonstrates the rich physics underlying CMHD.

\section{Summary and Implications}
We have presented a formulation of hydrodynamic theory for Chiral fluid with dynamical magnetic field based on a generalization of derivative expansion. 
We derive the manifestation of CME in CHMD at the first order in this expansion scheme. 
It would be interesting to extend our formulation to a higher order to study CVE and other anomaly-induced phenomena in CMHD (e.g. Ref.~\cite{Hattori:2016njk}). 
In addition, we explore the real-time properties of Chiral fluid, and find qualitative difference in the aspects of stability and polarization of collective modes.

In this work, we focus on the basic formulation and general properties of Chiral fluid. 
Our findings can be applied to specific Chiral matter, such as Weyl semimetal~\cite{PhysRevLett.121.176603} and QGP created in heavy-ion collisions. 
As for the later, the dynamics of baryon density can be potentially important, and is particular relevant to the coming low-beam-energy scan at RHIC. 
It would be interesting to extend the present analysis of CMHD by including $n_{B}$ as well.

\begin{acknowledgements}
This work is partially supported by the
U.S. Department of Energy, Office of Science, Office of Nuclear Physics,
with grant Nos.~DE-SC0012704 (Y.H.), DE-SC0018209 (H.-U.Y.), and DE-SC0011090 (Y.Y.), and within the framework of the Beam Energy Scan Theory Topical Collaboration. 
K.H. is supported by the China Postdoctoral Science Foundation
under grant Nos.~2016M590312 and 2017T100266. 
We thank Paolo Glorioso, Xu-Guang Huang, Matthias Kaminski, Dam Son and Naoki Yamamoto for insightful comments on the manuscript.
We are grateful to anonymous referees for insightful suggestions which improve the manuscript significantly.
\end{acknowledgements}


%
%
%
%

\begin{appendix}
\section{The constitutive relation for Chiral MHD}
\label{sec:cmhd}

In this section, 
we will supplement the discussion in Sec.~\ref{sec:CMHD} with more details on the form of the constitutive relation of CMHD. 
We shall see the second law of thermodynamics imposes important constraint on such relation.

We would like to express
$\{T^{\mu\nu}, J_{\rm A}^\mu, \tilde F^{\mu\nu}\}$ in terms of CMHD
variables $\{\e, B,n_{A}\}$ by generalizing the standard procedure of derivative
expansion to the double expansion in terms of gradient and $C_{A}$:
\begin{equation}
\label{delta-expand}
 T^{\mu\nu}  = T^{\mu\nu}_{(0)} +  T^{\mu\nu}_{(1)}\, ,
  \quad
 \tF^{\mu\nu}= \tF^{\mu\nu}_{(0)}+ \tF^{\mu\nu}_{(1)}\, ,
  \quad
  J_{\rm A}^\mu =  J_{\rm A(0)}^\mu + J_{\rm A(1)}^\mu, 
%
\end{equation}
as we explained earlier.
We will use subscript $(0), (1)$ to denote quantities at zeroth and first order in $\d$ respectively. 
As a reminder,
$\CO(\delta)$ denotes quantities of the order $\CO(k)$ and/or $\CO(C_{A})$. 
CMHD variables are counted as zeroth order in $\CO(\delta)$. 
Non-CMHD variables are counted as the first order or even the higher order in $\d$ .
In particular, $E^{\mu}$ is counted as $\CO(\d)$. 

At zeroth order in $\d$,
the (total) energy momentum tensor, the axial current and $\tF^{\mu\nu}$ can be written in general as 
\bes
\label{zeroth-order-constitute}
\begin{eqnarray}
T_{(0)}^{\mu\nu} &=& \epsilon \, u^\mu u^\nu - X \Delta^{\mu\nu}  
-Y B^\mu B^\nu\, , 
\\
\tF^{\mu\nu}_{(0)} &=& B^{\mu} u^{\nu} - B^{\nu}\, u^{\mu}\, ,
\\
J_{{\rm A}(0)}^\mu &=& n_{\rm A} u^\mu 
\,,
\end{eqnarray}
\ees
where 
\begin{eqnarray}
\Delta^{\mu\nu}  = g^{\mu\nu} - u^\mu u^\nu\, . 
\end{eqnarray}
Here, $X$, and $Y$ are functions of CMHD variables, as we shall determine shortly. 
A term proportional to $B^{\mu}$ is not allowed in $J^{\mu}_{\rm A(0)}$ based on the symmetry consideration. 
For such a term to be present, its prefactor has to be C-odd, since
$B^{\mu}$ is C-odd and $J^{\mu}_{\rm A}$ is C-even. 
However, the electric charge density, $n$ vanishes in the limit $\d\to 0$, and so is the prefactor of such term. 
For the same reason, we do not include a term proportional to
$(B^\mu u^\nu + B^\nu u^\mu)$ in $T^{\mu\nu}_{(0)}$. 
Note that the expression for $\tF^{\mu\nu}_{(0)}$ follows from the definition of $B^{\mu}$.

%
%


In what follows, we will obtain the expression for $X, Y$ and $T^{\mu\nu}_{(1)},J^{\mu}_{\rm A(1)}, E^{\mu}_{(1)}$ via the second law of thermodynamics.
Note by definition $E^{\mu}_{(1)}$ is related to $\tF^{\mu\nu}_{(1)}$ as:
\begin{eqnarray}
\label{tF-1}
\tF^{\mu\nu}_{(1)}
&=&
\epsilon^{\mu\nu\alpha\beta} u_\alpha E_{(1)\beta}\, . 
\end{eqnarray}
For this purpose, we consider the entropy current $S^\mu$:
\begin{equation}
 S^\mu = S^{\mu}_{(0)}+S^{\mu}_{(1)}\, ,
\end{equation}
where the zeroth order entropy current is given by
\begin{eqnarray}
S^{\mu}_{(0)}=s u^{\mu}\, . 
\end{eqnarray}
The divergence of $S^{\mu}$ now reads
\begin{eqnarray}
\label{ds-A}
 \partial_{\mu} S^{\mu}
 &=& D s + s \, \theta +\pd_{\mu}\,S^{\mu}_{(1)}
 \nonumber\\
 &=& \beta D \epsilon - \(\b \mu_{\rm A}\) D n_{\rm A} + \beta H_\mu D
 B^\mu + s \, \theta
 \nonumber\\ 
  &\,&+\pd_{\mu} S^{\mu}_{(1)}\, , 
\end{eqnarray}
where from the first line to the second line we have used \eq\eqref{ds}. 
Here we have introduced the short-handed notation:
\begin{eqnarray}
D \equiv u_\mu \p^\mu\, , 
\qquad
\theta \equiv \p_\mu  u^\mu\, . 
\end{eqnarray}

On the other hand, 
by substituting \eq\eqref{zeroth-order-constitute}, \eq\eqref{tF-1} and \eq\eqref{delta-expand} into \eqref{MHD-eq}, we have
\bes
\label{eq-expand}
\begin{eqnarray}
\label{De}
 D \epsilon  &=&- \(\epsilon+X\) \, \theta - Y B_\nu (B\cdot \p) u^\nu
\nonumber\\
&\,&
+ \( \p_\mu u_\nu\)\, T_{(1)}^{\mu\nu} \, , 
\\
\label{DB}
D B^\mu&=&
 -B^\mu \theta +(B \cdot \p) u^\mu + u^\mu (\p \cdot B)
\nonumber\\
&\,&+\pd_{\mu}\,  \left(
\epsilon^{\mu\nu\alpha\beta} u_\alpha E_{(1)\beta}
	   \right)
\\
\label{DnA}
D n_{A}&=& -n_{A}\,\theta
\nonumber\\
&\,&
-\pd_{\mu}\, J^{\mu}_{A (1)}- C_{\rm A} E_{(1)} \cdot B\, .
\end{eqnarray}
\ees
In \eq\eqref{De}, 
we have used the fact that $u_\nu T^{\mu\nu}_{(1)}=0$ since we are working in the Landau fluid frame throughout. 
Combining \eq\eqref{eq-expand} and \eq\eqref{ds-A} yields:
\begin{eqnarray}
\label{ds-Final}
 \partial_{\mu} S^{\mu}
 &=& D s + s \, \theta +\pd_{\mu}\,s^{\mu}_{(1)}
 \nonumber\\
 &=&
 \beta \theta (- X - \epsilon + sT + \mu_{\rm A}  n_{\rm A} - H \cdot B) 
\nonumber\\
&\,& + \beta  \(  H_\mu -  Y B_\mu \)\(B \cdot \p\)u^\mu 
\nonumber\\
&\,&  +\beta\,\( \p_\mu u_\nu\)  T^{\mu\nu}_{(1)} 
 + \(\beta \mu_{\rm A}\) \left(  \p \cdot J_{\rm A(1)} + C_{\rm A} E_{(1)} \cdot B
 \right)
\nonumber\\
&\,& 
+ \beta H_{\mu}
\p_\nu \left(
\epsilon^{\mu\nu\alpha\rho} u_\alpha E_{(1)\rho}
 \right)
 + \p \cdot S_{(1)} \, . 
\end{eqnarray}
Eq.~\eqref{ds-Final} will be the starting point of the analysis in subsequent sections~\ref{sec:ideal}, \ref{sec:first-order}.

\subsection{The constitutive relation at zeroth order in $\delta$: ideal chiral MHD
\label{sec:ideal}
}

To determine $T^{\mu\nu}_{(0)}$, we consider \eq\eqref{ds-Final} at first order in $\d$. 
By requiring that there is no entropy production at this order, we have: 
\begin{eqnarray}
\label{ds-zero}
\pd\cdot S_{(0)}&=&
 \beta \theta (- X - \epsilon + sT + \mu_{\rm A}  n_{\rm A} - H \cdot B) 
  \nonumber\\
&\, &
-  \beta  \(  H_\mu -  Y B_\mu \)(B \cdot \p)u^\mu  
\nonumber\\
&=&0\, . 
\end{eqnarray}
Therefore, $X$, $Y$ satisfy 
\bes
\label{XY}
\begin{eqnarray}
\label{Y}
 H_\mu&=& Y B_\mu\, ,
 \\
 \label{X}
X &=& -\epsilon + sT + \mu_{\rm A}  n_{\rm A} - H \cdot B  \, , 
\end{eqnarray}
\ees
meaning $Y B^{\mu}$ is the in-medium magnetic field $H^{\mu}$ and $X$ must be identified with pressure $p$ after comparing \eq\eqref{X} with \eq\eqref{p}. 
Substituting \eq\eqref{XY} into \eq\eqref{zeroth-order-constitute}, 
we obtain $T^{\mu\nu}_{(0)}$ used in \eq\eqref{eq:constitutive}:
\begin{eqnarray}
\label{Tmunu-zero}
T^{\mu\nu}_{(0)}
=
\e\, u^{\mu}u^{\nu} - p \Delta^{\mu\nu}- H^{\mu}B^{\nu}\, . 
\end{eqnarray}

While in a large body of literature on relativistic MHD, energy density and pressure are often divided into the fluid part and EM part (c.f.~Refs.~\cite{Huang:2009ue,Huang:2011dc}), 
in this work, $\e$ and $p$ denote total energy and pressure respectively, including the contributions from both fluid and EM field. 
The reason for doing so is that making the above-mentioned division could be physically impossible if the constitute of the fluid is strongly coupled to the EM field. 
As commented by
W. Israel in Ref.~\cite{Israel1978}: ``questions about which part should be called the 'electromagnetic energy tensor' are near semantics and to a large extent, superfluous''. 
That said, 
we could recover the conventional constitutive relation of  ideal MHD from \eq\eqref{Tmunu-zero} as follows. 
To simplify the discussion, we will assume $H^{\mu}=B^{\mu}$, 
and decompose the total energy density and pressure in the following form, 
\begin{equation}
 \epsilon = \epsilon_{f} + \frac{1}{2} \CB^2,
  \quad
 p = p_{f}+\frac{1}{2} \CB^2\, , 
\end{equation}
with $e_{f}, p_{f}$ being the fluid parts
and $\CB=\sqrt{-B\cdot B}$. 
Then energy momentum tensor of the conventional ideal MHD (c.f.~Eq.~(3) and (4) in Ref.~\cite{Huang:2011dc}) is reproduced as 
\begin{eqnarray}
 T^{\mu\nu}_{(0)}
=T^{\mu\nu}_{(0){\rm fluid}}+T^{\mu\nu}_{(0){\rm EM}}\, , 
\end{eqnarray}
where
\begin{eqnarray}
 T^{\mu\nu}_{(0){\rm fluid}} &=&
  \epsilon_{f}  u^\mu u^\nu
  -  p_{f}  \Delta^{\mu\nu}
 \\
 T^{\mu\nu}_{(0){\rm EM}} &=&  
\CB^{2}\, u^{\mu}u^{\nu}-\frac{1}{2}\CB^2 g^{\mu\nu} - B^\mu B^\nu
   \,. 
\end{eqnarray}
We note $ T^{\mu\nu}_{(0){\rm EM}}$ is the conventional EM stress-energy tensor in the absence of electric field.

\subsection{The constitute relation at first order in chiral MHD and CME
\label{sec:first-order}
}

To constrain the constitutive relation at $\CO(\d)$, we consider \eq\eqref{ds-Final} at second order in $\d^{2}$. 
The entropy production has to be positive-definite at this order. 
Therefore:
\bes
\begin{eqnarray}
&\,&
\beta\,\( \p_\mu u_\nu\)  T^{\mu\nu}_{(1)} 	
 + \beta \mu_{\rm A} \left(  \p \cdot J_{\rm A(1)} + C_{\rm A} E_{(1)} \cdot B
 \right)
\nonumber\\
&\,& 
+ \beta H_{\mu}
\p_\nu \left(
\epsilon^{\mu\nu\alpha\beta} u_\alpha E_{(1)\beta}
 \right)
 + \p \cdot S_{(1)} \, . 
\nonumber\\
&=&
\partial_{\mu}\[ S^{\mu}_{(1)}
 + \beta \mu_{\rm A} J_{\rm A(1)}^\mu
-
\epsilon^{\mu\nu\alpha\beta} \beta H_{\nu} u_\alpha E_{(1)\beta}   \]
\label{s1}
\\
&\,&
+T^{\mu\nu}_{(1)}\p_\mu (\beta  u_\nu)
 - J_{\rm A(1)} \cdot \p \(\b \mu_{\rm A}\)
\nonumber \\
&\,&+ E_{(1)\rho}
 \[
  C_{\rm A} \beta \mu_{\rm A} B^{\rho}
 +
 \epsilon^{\mu\nu\alpha\rho}
 \p_\mu (\beta H_{\nu})
 u_\alpha
 \] \,, 
 \nonumber\\
 &\geq& 0\, ,
\end{eqnarray}
\ees

 We first identify the first-order entropy current as \eq\eqref{s1}:
\begin{equation}
\label{s1-result}
 s_{(1)}^\mu
  =
 - \beta \mu_{\rm A} J_{\rm A(1)}^\mu
+
\epsilon^{\mu\nu\alpha\rho} \beta H_{\nu} u_\alpha E_{(1)\rho} \, . 
\end{equation}
Furthermore, 
the following relations are sufficient to guarantee the positivity of entropy production
\bes
\label{condition-detail}
\begin{eqnarray}
&&
 T^{\mu\nu}_{(1)}\(\p_\mu u_\nu\) \ge 0 \,,
  \label{eq:t1-cons}
  \\
&&
 - J_{\rm A(1)} \cdot \p\( \beta \mu_{\rm A}\) \ge 0\, , 
  \label{eq:j1-cons}
  \\
&&
  E_{(1)\rho}
 \[
  C_{\rm A} \beta \mu_{\rm A} B^{\rho}
 +
 \epsilon^{\mu\nu\alpha\rho}
 \p_\mu (\beta H_{\nu})
 u_\alpha
 \] 
\ge 0\, .
\label{eq:e1-cons}
\end{eqnarray}
\ees
Equations (\ref{eq:t1-cons}, \ref{eq:j1-cons}) are satisfied if we
introduce the viscosities and axial conductivity as 
\begin{eqnarray}
 && T^{\mu\nu}_{(1)} =2 \eta \nabla^{< \mu} u^{\nu > }_{\perp}
  + \zeta \Delta^{\mu\nu} \nabla_{\perp} \cdot
  u\, , 
  \\
&& J^{\mu}_{\rm A(1)} = \l_{\rm A} \nabla^{\mu}_{\perp} \(\beta \mu_{\rm A}\)\,,
\end{eqnarray}
where $\<\ldots\>$ denotes the symmetric and traceless part, 
and $\nabla^\mu_{\perp} \equiv \Delta^{\mu\nu}\p_\nu$. 
In general, the viscosities and the diffusion constant $\l_{A}$ can be
anisotropic, due to the existence of the magnetic field 
(see Refs.~\cite{Huang:2009ue,Huang:2011dc, Kovtun:2016lfw,Grozdanov:2016tdf,Hernandez:2017mch} for further discussions
and Refs.~\cite{Tuchin:2011jw,Fukushima:2015wck,Li:2017tgi,Finazzo:2016mhm,Hattori:2017qih} for recent computations of those anisotropic transport coefficients for hot QCD matter). 

We now focus on Eq. (\ref{eq:e1-cons}) which implies 
\begin{equation}
\[
  C_{\rm A} \beta \mu_{\rm A} B^{\rho}
 +
 \epsilon^{\mu\nu\alpha\rho}
 \p_\mu (\beta H_{\nu})
 u_\alpha
 \] = Z^{\rho\nu} E_{(1)\nu} \,,
\end{equation}
with a semi-positive definite $Z^{\mu\nu}$. 
To further constrain the form of $Z^{\mu\nu}$, 
it is convenient to introduce a unit vector in the direction of the magnetic field, 
\begin{equation}
 b^{\mu} \equiv \frac{ B^\mu }{ \sqrt{ |B|}}, 
\end{equation}
which satisfies $b_\mu b^{\mu} = -1$.
The spatial projector can be decomposed into the directions parallel and
perpendicular to $\bm B$, as 
\begin{equation}
\Delta^{\mu\nu} = - b^{\mu}b^{\nu} + \Delta_{\perp}^{\mu\nu}\,,
\end{equation}
where $\Delta_{\perp}^{\mu\nu} \equiv  \Delta^{\mu\nu} + b^{\mu}b^{\nu}$. 
We can introduce two independent dissipative conductivities
and the Hall conductivity consistent with the second law of thermodynamics as  
\begin{equation}
Z^{\mu\nu} =
  - \beta \left[
	   - \sigma_{||} b^{\mu}b^{\nu}
	   + \sigma_{\perp} \Delta_{\perp}^{\mu\nu}
	+ \sigma_{\rm Hall}\, \epsilon^{\mu\nu\alpha\beta}u_\alpha b_\beta
   \right]\, .
\end{equation}
The reader is referred to Refs.~\cite{Hattori:2016lqx,Hattori:2016cnt,Fukushima:2017lvb} for recent computation of $\s_{\parallel}, \s_{\perp}$ for weakly coupled QGP. 
Note that the term involving the Hall conductivity does not produce
entropy. 
Let us look at the limit where  $\sigma_{||} = \sigma_{\perp}\equiv \sigma$ and
$\sigma_{\rm Hall}=0$, in which case 
\begin{equation}
\[
  C_{\rm A}  \mu_{\rm A} B^{\rho}
 +
 \epsilon^{\mu\nu\alpha\rho}
 \b^{-1}\, \p_\mu (\beta H_{\nu})
 u_\alpha
 \]
 =- \s E_{(1)}^\rho\, . 
\label{eq:e1}
\end{equation}
From \eq\eqref{eq:e1}, we obtain \eq\eqref{E-relation}. 
The second term on the right in Eq. (\ref{eq:e1}) 
represents the CME, 
as we explained in Sec.~\ref{sec:CMHD}.

\section{A discussion on vector charge density and electric filed $E^{\mu}$}
\label{sec:nvandE}

In this work, we did not include vector charge density $n$ and electric field $F^{\mu}$ as CMHD variables. 
As we explained in Introduction.~\ref{sec:intro}, 
both $n$ and $E^{\mu}$ are damped out at rate proportional to the electric conductivity $\s$.
The reason can be understood as follows.
In this paragraph, we will discuss the damping of $n$. 
The case for electric field is similar. 
To see the key physics in the simplest way, let us consider the non-relativistic case.
If we substitute the Ohm's law $\bm j = \sigma \bm E$ for the current
conservation, and use the Gauss law $\nabla \cdot \bm E = n$, we obtain
\begin{equation}
 \p_t n = - \sigma n , 
\end{equation}
which indicates that $n$ is damped out with the rate $\sigma$.
Instead, the value of $n$ is slaved by the hydrodynamic variables. 
For example in ref.~\cite{Hattori:2016njk}, it is found that 
$n=C_{A}\o\cdot B$ for a chiral fluid in strong magnetic field limit.

We note in passing that the relativistic Ampere's law:
\begin{eqnarray}
\label{Ampere}
 \partial_\mu F^{\mu\nu}=J^\nu\, , 
\end{eqnarray}
can be recasted into a relaxation equation which tells tush's hat $E^{\mu}$ will approach the expression given by \eq\eqref{E-relation} at time scale much longer than $1/\s$. 
To make the argument simple and clear, let us consider a local rest frame where \eqref{Ampere} takes the usual form in terms of spatial electric and magnetic fields
\be
-{\partial\vE\over \partial t}+{\bm\nabla}\times \vB={\bm
J}=\sigma\vE+C_{\rm A}\mu_{\rm A}\vB\, ,
\ee
where we have used \eq\eqref{jv}.
This may be written as 
\be
{\partial \vE\over\partial
t}=-\sigma\[\vE-\({1\over\sigma}{\bm\nabla}\times \vB+{C_{\rm
A}\mu_{\rm A}\over\sigma}\vB\)\]\,.\label{israel-stewart}
\ee
What appears inside the bracket in the right-hand side is precisely the first order constitutive relation for $\vE$ (c.f.~ \eq\eqref{E-rest}), and the above equation 
is the relaxation equation of $\vE$ to its constitutive relation with the relaxation time $1/\sigma$.
This is similar to the Israel-Stewart theory of dissipative hydrodynamics.
In the CMHD time scale ( $\gg 1/\sigma$), the iterations of
(\ref{israel-stewart}) will reduce to the conventional derivative
expansions in space-time and $C_{\rm A}$ that our study is based on, but the microscopic theory is consistent with causality
in a way similar to how the Israel-Stewart theory restores causality.
This suggests that a numerical simulation of CMHD that is consistent with causality may need to use the original Maxwell equations {\it a la} Israel-Stewart theory, instead of a finite truncation of derivative expansions.
%
%



\section{A Kubo formula for chiral magnetic conductivity in CMHD}
In the literature, 
$\s_{\rm A}\equiv C_{\rm A}\, \mu_{A}$ is sometimes referred to 
as the chiral magnetic conductivity~\cite{Kharzeev:2009pj}. 
It is useful to derive a Kubo formula for $\s_{\rm A}$ in CMHD. 
For this purpose, we consider a fluid at rest, i.e $u^{\mu}=\(1,0,0,0\)$
and replace $\vj$ in \eq\eqref{jv} with $ \nabla\times \vH$.
Keeping only contribution from CME in \eq\eqref{jv}, 
we obtain:
$\vB = \s^{-1}_{\rm A}\, \nabla\times \vH$.
Since $\vH$ is conjugate to $\vB$, we then have:
\begin{equation}
\label{sigmaA-Kubo}
\s^{-1}_{\rm A} =\lim_{\vk\to 0\,\o\to 0} \frac{1}{k_{l}}\, \e_{ijk}\<\,B^{i}\, B^{j}\,\>\, ,
\end{equation}
where $\<\,B^{i}\, B^{j}\,\>$ denotes the retarded Green's function of $\vB$ in Fourier space. 
\eqref{sigmaA-Kubo} is different from Kubo formula for $\s_{\rm A}$ in usual cases where magnetic field is considered to be non-dynamical (c.f.~\cite{Kharzeev:2009pj}). 
Here $\s_{\rm A}$ is related to the retarded Green's function of the magnetic field.
The Kubo formula \eq\eqref{sigmaA-Kubo}, which is new in the literature, 
opens a possibility to compute $\s_{\rm A}$ in chiral systems in which EM fields are dynamical.

\section{Analytic solutions of the secular equation in \eq\eqref{VG-eq} with limiting values of $\theta$
\label{sec:sol}
}

When $ \theta = 0 $, \eq\eqref{VG-eq} reduces to  
\begin{eqnarray}
 (\VgSquare - \tilde c_{\rm s}^2 )  \[ \,  (\VgSquare - 1)^2 + \tilde
  \epsilon_{\rm A}^2 \VgSquare  \, \] = 0
 \, ,
\end{eqnarray}
where $\tilde{\e}_{\rm A}=\e_{\rm A}/u_{\rm A}, \VgSquare=V/u_{A}$, and we find the six solutions as 
\begin{eqnarray}
V = \pm c_{\rm s}  , \ \ 
 V = \pm \left( \sqrt{u_{\rm A}^2 - \frac{\epsilon_{\rm A}^2}{4} }
	  \pm i \frac{\epsilon_{\rm A}}{2} \right) 
\, ,
\end{eqnarray}
where the signs are taken for all combinations. 
The first two are ordinary sound modes without any modification by the magnetic field or anomaly, 
since the pressure of the magnetic field does not contribution to that along the wave vector. 
The remaining four modes arise as the results of the mixing between the \Alf and magnetosonic waves, 
where the $  V$ are complex when $\epsilon_{\rm A}^2 < 4 u_{\rm A}^2  $, 
and are pure imaginary when $\epsilon_{\rm A}^2 > 4 u_{\rm A}^2  $. 
Those regions correspond to Phase IIA and Phase IIB, respectively. 
In the both phases, positive and negative imaginary parts appear in pairs, 
and the signs depend on that of $ \epsilon_{\rm A} $.

When $ \theta = \pm \pi/2 $, the secular equation reduces to  
\begin{eqnarray}
 \VgSquare \left[ \,
	    \VgSquare^2
	    - \VgSquare (  1 + \tilde c_{\rm s}^2 - \tilde \epsilon_{\rm A}^2)
	    - (\tilde \epsilon_{\rm A}  \tilde c_{\rm s} )^2 \, \right] = 0 
\, ,
\end{eqnarray}
and we find the six solutions as 
\begin{eqnarray}
V = 0  , \ \ 
V = \pm \sqrt{ \kappa \pm \sqrt{ \kappa^2 + \nu }  }
\, ,
\end{eqnarray}
where $  \kappa =  ( u_{\rm A}^2 + c_{\rm s}^2 - \epsilon_{\rm A}^2) $
and $ \nu =  (\epsilon_{\rm A} c_{\rm s} )^2  $.
The presence of two vanishing solutions means group velocity becomes zero.
There are two real solutions, which provide the waves propagating in the opposite directions. 
Their velocities are modified by the anomaly effect. 
The remaining two solutions are vanishing in the absence of anomaly effects ($ \nu =0 $), 
which, however, become a pair of positive and negative pure imaginary numbers when $ \nu \not =0 $. 
They do not propagate, but grows or dissipates exponentially in time, respectively. 

\end{appendix}

\bibliography{CMHD}

\end{document}